%% file: main.tex
\documentclass[conference]{IEEEtran}
\IEEEoverridecommandlockouts

\input{packages}

\input{notation}

\def\BibTeX{{\rm B\kern-.05em{\sc i\kern-.025em b}\kern-.08em
    T\kern-.1667em\lower.7ex\hbox{E}\kern-.125emX}}
\begin{document}

\title{How Inclusive Are Wikipedia's Hyperlinks in Articles Covering Polarizing Topics?}

\author{\IEEEauthorblockN{Cristina Menghini}
\IEEEauthorblockA{\textit{Computer Science} \\
\textit{Brown University}\\
Providence, Rhode Island, USA  \\
cristina\_menghini@brown.edu}
\and
\IEEEauthorblockN{Aris Anagnostopoulos}
\IEEEauthorblockA{
\textit{Sapienza University}\\
Rome, Italy \\
aris@diag.uniroma1.it}
\and
\IEEEauthorblockN{Eli Upfal}
\IEEEauthorblockA{\textit{Computer Science} \\
\textit{Brown University}\\
Providence, Rhode Island, USA  \\
eli\_upfal@brown.edu}
}

\maketitle

\input{abstract}


\input{introduction}

\input{related}
\input{prelims}

\input{metrics}

\input{exposure}
\input{concl}






\bibliographystyle{IEEEtran}
\bibliography{IEEEabrv,poltopics}

\end{document}

%% file: packages.tex
\usepackage{amsmath}
\usepackage[utf8]{inputenc}
\usepackage{booktabs}
\usepackage{caption}
\usepackage{color,soul}
\usepackage{xspace}
\usepackage{nicefrac}
\usepackage{cite}
\usepackage{amsmath,amssymb,amsfonts}
\usepackage{algorithmic}
\usepackage{textcomp}
\usepackage{xcolor}
\usepackage{hyperref}
\usepackage{cleveref}

\Crefname{section}{Section}{Sections}
\crefname{section}{Sect.}{Sect.}
\crefname{figure}{Fig.}{Fig.}
\Crefname{figure}{Figure}{Figure}
\crefname{table}{Tab.}{Tab.}
\Crefname{table}{Table}{Table}
\crefname{definition}{Def.}{Def.}
\Crefname{definition}{Definition}{Definition}
\usepackage[multiple]{footmisc}
\usepackage{graphicx} 
\usepackage{graphicx}  
\usepackage{multirow}
\usepackage{subfigure}
\usepackage{multicol}
\usepackage{ifthen}
\usepackage{mathtools}
\newboolean{shortver}
\setboolean{shortver}{true}

\usepackage[show]{chato-notes}

%% file: notation.tex
\newtheorem{definition}{Definition}

\newcommand{\topinet}{topic-induced network\xspace}
\newcommand{\topinets}{topic-induced networks\xspace}
\newcommand{\Adivexp}{ExDIN\xspace} 
\newcommand{\Amutual}{M-ExDIN\xspace} 
\newcommand{\divexp}{exposure to diverse information\xspace}
\newcommand{\udivexp}{Exposure to diverse information\xspace}
\newcommand{\mutual}{Mutual exposure to diverse information\xspace}
\newcommand{\lmutual}{mutual exposure to diverse information\xspace}

\newcommand{\topicpages}{\ensuremath{\mathcal{T}}\xspace} 
\newcommand{\wikipages}{A} 
\newcommand{\wikiedges}{L} 
\newcommand{\p}{P} 
\newcommand{\pbar}{\bar{P}}

\newcommand{\outedges}[1]{N_{out}(#1)}
\newcommand{\onehop}{\mathcal{N}}

\newcommand{\exposure}[3]{e_{#1 \rightarrow #2}^#3} 

\newcommand{\specialcell}[2][c]{%
  \begin{tabular}[#1]{@{}c@{}}#2\end{tabular}}

%% file: abstract.tex
\begin{abstract}
\emph{Wikipedia} relies on an extensive review process to verify that
the content of each individual page is unbiased and presents a ``neutral
point of view.'' Less attention has been paid to possible biases in the hyperlink structure of Wikipedia, which has a significant influence on the user's exploration process when visiting more than one page.
The evaluation of hyperlink bias is challenging because it depends on the global view rather than the text of individual pages.

In this paper, we focus on the influence of the interconnect topology between articles describing complementary aspects of polarizing topics.
We introduce a novel measure of \emph{exposure to
diverse information} to quantify users' exposure to
different aspects of a topic throughout an entire surfing session, rather than just one click ahead.
We apply this measure to six polarizing topics (e.g., \emph{gun control} and
\emph{gun right}), and we identify cases in which the network topology significantly limits the exposure of users to diverse information on the topic, 
encouraging users to remain in a
\emph{knowledge bubble}. 
Our findings demonstrate the importance of evaluating Wikipedia's network structure in addition to the extensive review of individual articles.

\end{abstract}

%% file: introduction.tex
\section{Introduction}\label{sec:intro}
Knowledge on Wikipedia is distributed across articles inter-connected via hyperlinks.
According to Wikipedia's Linking Manual~\cite{wikipediaManualStyle}, \emph{``Internal links can add to the cohesion and utility of Wikipedia, allowing readers to deepen their understanding of a topic by conveniently accessing other articles.''}
Consequently, users are \emph{directly} exposed to an article's content and \emph{indirectly} exposed to the content of the pages it points to. 

Wikipedia’s pages offer high-quality content with emphasis on an unbiased, neutral point of view (NPOV)~\cite{keegan2011hot,piscopo2019we,wikipediaNPOV}, thanks to  numerous policies and guidelines~\cite{beschastnikh2008wikipedian,forte2009decentralization}.
Although it provides tools to support the community for curating pages, it lacks a systematic way to contextualize them within the more general articles' network. 
Indeed, it is hard to evaluate the extent to which the current hyperlinks satisfy their purpose, especially in connecting articles related to a broad topic. 

The majority of users who look for a specific information are likely to find their answers on the first Wikipedia page they are visiting~\cite{singer2017we},
whereas about $20\%$ of Wikipedia's users follow hyperlinks within
Wikipedia to develop a broad view of a subject.\footnote{For the English Wikipedia, the number of unique devices is around 800 million per month.
If a device corresponds to a user, around 160 million of users click at least one link throughout their visit on Wikipedia~\cite{wikipediaStats}.}
It is therefore important to investigate whether the link structure leads users to visit pages presenting broad and diverse aspects of their topic of interest. 
We initiate this important study by concentrating on polarizing topics spanning across multiple articles.

For example, consider the topic \emph{abortion}, which is distributed across multiple articles on Wikipedia. 
Because of its \emph{polarizing} nature, we recognize pages about
events, people, or organizations that are associated either with \emph{pro-choice} or \emph{pro-life}. 
For instance, the page \emph{Abortion-rights movements} describes organizations related to \emph{pro-choice} view. 
In this particular page, we identify 15 links pointing to articles about \emph{pro-choice} subjects and only 3 hyperlinks directed to  \emph{pro-life} related pages. 
Furthermore, if we consider articles at distance 2 from the page \emph{Abortion-rights movements}, then
there are 4 times more pages associated with  \emph{pro-choice} than articles associated with
\emph{pro-life} subjects.
Similar counting, starting  from the page \emph{Anti-abortion movements}, shows 18 outgoing links to
\emph{pro-life} pages and only 1 to a \emph{pro-choice} article. 
At distance two we have 15 times more pages related to the category \emph{pro-life} than pages related to \emph{pro-choice}.

The example above demonstrates unbalanced hyperlink structure in Wikipedia that may influence users' exposure to diverse information on a topic. 
On another, albeit very different, platform
a recent work~\cite{ribeiro2020auditing} empirically showed that its recommender system contributes to radicalizing users' pathways.
Given the major role of Wikipedia as a popular primary source of knowledge, it is important to evaluate the effect of its hyperlink structure on user navigation, to guarantee a balanced access to well-rounded knowledge. 

Evaluating the influence of the hyperlink topology is challenging because it
requires a broad view of the network topology, not just the text of a single article. Such a view, and a technique to analyzing it, is not readily available to Wikipedia editors. In this work, we develop an algorithmic approach to quantify users' exposure among a set of articles. Then, we audit the extent to which the current Wikipedia's link structure allows users to browse different stances of polarizing topics. 

Our main contributions are the following:
\begin{itemize}
    \item We initiate the study of the hyperlink network's role in driving users to explore articles of different categories. 
    We investigate this on a set of polarizing topics.
    
    \item We design two metrics, the \emph{\divexp} and the
    \emph{(mutual)} \emph{\divexp}, which quantify the likelihood of visiting pages belonging to different sets of articles click-after-click (\cref{sec:metrics}).

    \item We apply our new measures to Wikipedia hyperlink sub-graphs related to six polarizing topics. 
    We identify cases in which the Wikipedia's hyperlink network significantly limits the exposure of users to diverse information on the topic (\cref{sec:exposure_main}).
\end{itemize}

The code to replicate the analysis is available at \url{https://github.com/CriMenghini/WikiNetBias}.

%% file: related.tex
\section{Related Works}\label{sec:relwork}

\textbf{\emph{Improving Wikipedia}}. 
Previous works proposed semi-automated procedures to improve Wikipedia's
quality by checking the veracity of references
\cite{redi2019citation,fetahu2016finding}, suggesting articles'
structure \cite{piccardi2018structuring}, looking for hoaxes
\cite{oaxes}, or recommending links~\cite{paranjape2016improving, wulczyn2016growing}.
Among these tools, none provides a measure to evaluate the link-based relationship across articles of diverse categories.
In this work, we define such metrics (\cref{sec:exposure}).

\textbf{\emph{Wikipedia Navigation}}. 
The literature still lacks a model that generalizes Wikipedia's users' behavior.
Previous studies ~\cite{helic2013models, gildersleve2018inspiration,lamprecht2017structure,singer2013human} focused on modeling and predicting human navigation relying on traces from games~\cite{west2012human,scaria2014last,dallmann2016extracting,koopmann2019right,singer2013human}. 
Even though such games provide valuable insights on how users exploit links to move across concepts, other studies showed that users display different behavioral patterns depending on their information needs and the links' position within pages~\cite{singer2017we,dimitrov2016visual,dimitrov2017success}. 
We exploit such insights to define a general model mimicking localized and in-depth topic exploration (\cref{sec:metrics}). 

\textbf{\emph{Wikipedia Categorization}}. 
When dealing with polarizing topics, one needs to distinguish between pages belonging to different side of the topic. 
Because of the topic granularity, it is hard to rely on automated techniques to categorize articles. Thus, we refrain from using supervised tools as ORES\footnote{https://www.mediawiki.org/wiki/ORES} or topic modeling~\cite{blei2010probabilistic}, in favor of a mining procedure employed in~\cite{shi2019wisdom}, exploiting actual Wikipedia's categories (\cref{sec:data}).

\textbf{\emph{Polarization on Social Media}}. Many works aim to quantify polarization on social media~\cite{adamic2005political,CossardDFMKMPS20,conover2011political,flaxman2016filter,guerra2013measure,garimella2018quantifying,repbublik}. 
Random walk controversy~\cite{garimella2018quantifying} quantifies to what extent opinionated users are exposed to their own opinion rather than the opposite.
Bubble radius~\cite{repbublik} works on bipartite information networks and estimates the expected number of clicks to navigate from a page $v$ to any page of opposing opinion.
We focus on the metrics that better relate to our metric of \textit{\divexp} (\Adivexp). 
Differently from these two metrics, our measure of \textit{\divexp} works on multi-partite information networks and quantifies users' exposure to diverse information click-after-click. 

\textbf{\emph{Cultural bias on Wikipedia}}. Recent works found the
presence of cultural bias in the same articles of different languages \cite{callahan2011cultural} and gender biases \cite{genderbias,wagner2016women}. These content-based analyses prove that Wikipedia can be subjected to bias.
We decided to investigate bias on a novel topological perspective.

%% file: prelims.tex
\section{Preliminaries}\label{sec:data}

\begin{figure}[t]
\centering
\includegraphics[width=0.45\textwidth]{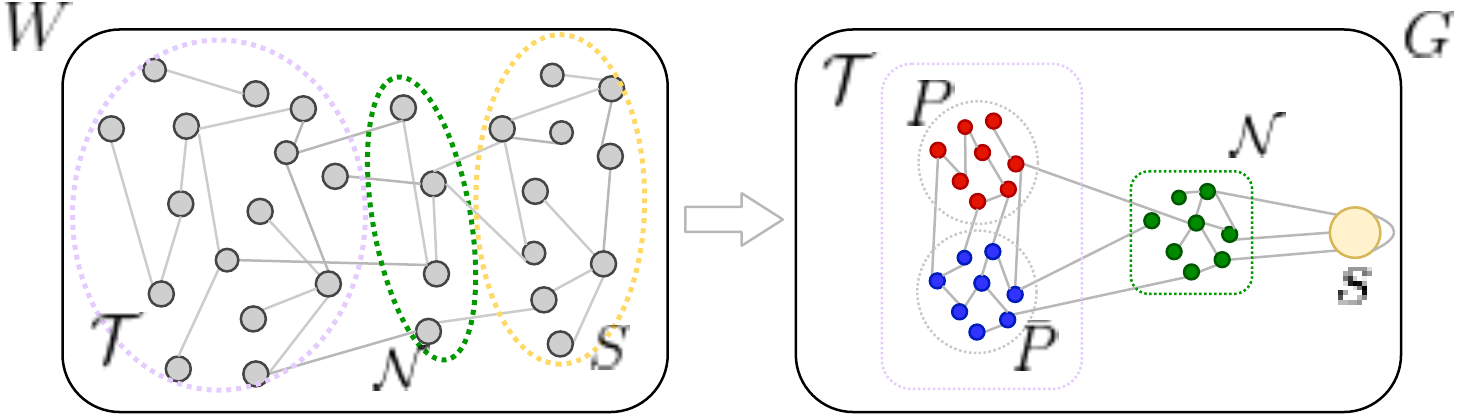}
\caption{On the left, the original Wikipedia graph. 
On the right, the final \topinet. 
The dashed circles in $W$ are the set of nodes used to build the \topinet\ $G$.
The colors red and blue refer to the sets $\p$ and $\pbar$, respectively. Green and yellow are $\onehop$ and $s$ respectively. 
We keep the image tidy and do not specify the edges' direction.}
\label{fig:graph}
\end{figure}

We encode a topic into a \topinet, a subgraph of 
the entire English Wikipedia's graph $W=(\wikipages,\wikiedges)$ (\cref{fig:graph}). 
The nodes of the graph are \emph{articles}~\cite{wikipediaNamespace},
and edges are links connecting pages, \emph{wikilinks}.\footnote{We
  exclude links within the same page and resolve all the
    redirects~\cite{wikipediaRedirect}. We do consider links in the
    infoboxes, which are summary standardized tables at the top--right corner of articles.}
Among all articles, we identify a set of pages $\topicpages \subset A$, about the topic. 
We partition these pages into two sets $\p$ and $\pbar$ (i.e., $\p \cap \pbar =
\emptyset$ and $\p \cup \pbar = \topicpages$), each gathering articles about the same side of the topic. 
In addition to the articles in $\topicpages$, we collect in $\onehop$ the pages at one-hop distance from them.
In this way, we can consider the chances of moving across partitions via articles not necessarily related to the topic.
To reduce the complexity of our analysis, we cluster all the pages in
$\wikipages \setminus (\topicpages \cup \onehop)$, into one super node $s$.
Note that $s$ is only connected to vertices in $\onehop$.
For each node $v\in\onehop$, we can have multiple edges going to $s$,
which we compress into one.
Respectively, $s$ can have multiple links to node $v \in
\onehop$, compressed into one as well. 

A \topinet is the directed weighted graph $G=(V,E)$, whose set of articles $V$ is $\topicpages \cup \onehop \cup \{s\}$ of size $n+1$, and edges $E$ are the links connecting them. 
The edge weights are \emph{transition probabilities} stored in a
row-stochastic matrix $M\in[0,1]^{\lvert V\rvert \times \lvert V\rvert}$, whose entry $m_{i,j}$ is the probability that from page $i$ a reader moves to $j$, and it is set to 0  if $(i,j)\not\in E$. 

In practice, to build a \topinet, we first extract the entire
Wikipedia's network from a complete English Wikipedia dump.\footnote{We refer to the dump of September 2020.}
Then, we only retain the graph induced by $\topicpages$, whose articles are selected and partitioned according to the strategy adopted in~\cite{shi2019wisdom}.
For instance, the topic \emph{abortion} polarizes into \emph{pro-life} ($\p$) and \emph{pro-choice} ($\pbar$) articles. 
The pro-life subcorpus consists of all articles categorized under the \emph{seed} category ``Anti-abortion movement'' and its subcategories. 
Similarly, we obtain the pro-choice corpus starting from the category
``Abortion-rights movement.''
Because we want the partitions to be disjoint, articles belonging to both ``Anti-abortion movement'' and ``Abortion-rights movement'' are assigned to~$\onehop$. 

In fact, as a consequence of Wikipedia's Neutral Point of View (NPOV)
policy~\cite{wikipediaNPOV}, we assume articles' content to
\emph{``fairly and proportionately represent all the significant views
that have been published by reliable sources on the topic.''} 
Moreover, as subcategories are often redundant or not entirely related
to the parent category, we check them manually, discarding categories whose names do not include topic-specific keywords.

\begin{table}[t]
\centering
\resizebox{\columnwidth}{!}{
\begin{tabular}{lcccc}  
\toprule
\textbf{Topic}    & \textbf{$\p$} & \textbf{$\pbar$} &  \textbf{Seed $\p$ }& \textbf{Seed $\pbar$} \\
\midrule
\textbf{\emph{Abortion}}       & Pro-life     & Pro-choice  & \specialcell{Anti-abortion\\ movement}     & \specialcell{Abortion-rights\\ movement}\\ \hline
\textbf{\emph{Cannabis}}       & Prohibition     & Activism & Cannabis prohibition     & Cannabis activism  \\ \hline
\textbf{\emph{Guns} }      & Control     & Rights  & \specialcell{Gun control \\ advocacy groups}     & \specialcell{Gun rights \\ advocacy groups}\\ \hline
\textbf{\emph{Evolution}}      & Creationism    & \specialcell{Evolutionary \\ biology} & Creationism    & \specialcell{Evolutionary \\ biology}\\ \hline
\textbf{\emph{Racism}}      & Racism   &  Anti-racism & Racism   &  Anti-racism\\ \hline
\textbf{\emph{LGBT}}      & Discrimination   &  Support & \specialcell{Discrimination against\\ LGBT people}   &  \specialcell{LGBT rights \\ movement} \\
\end{tabular}}
\caption{For each topic, the table indicates the partitions  $\p$ and
  $\pbar$ to which each standing corresponds. Moreover, we report the seed category for each partition.}
\label{tab:matching_partitions}
\end{table}

\subsection{Topic-Induced Networks}\label{sec:statistics}

We collect the \topinets\ related to six different polarizing topics: \emph{abortion, cannabis, guns, evolution, LGBT}, and \emph{racism} (\cref{tab:matching_partitions}). 

\begin{table}[t]
\centering
\resizebox{\linewidth}{!}{
\begin{tabular}{lccccccccc}  
\toprule
Topic    & $\vert V\setminus\{s\} \vert$ & $\vert \p \vert$ & $\vert \pbar \vert$ & $\vert \onehop \vert$  & $\vert E \vert$ & $\vert E_{\p \rightarrow \pbar} \vert$ &  $\vert E_{\pbar \rightarrow \p} \vert$ &  $\vert E_{\onehop \rightarrow \p} \vert$ & $\vert E_{\onehop \rightarrow \pbar} \vert$ \\ 
\midrule
\emph{Abortion}       & 56056     & 469  &  291  & 55296 & 2.1M & 205 & 97 & 21396 & 29889\\
\emph{Cannabis}       & 32743     & 45  &  231  & 32470 & 1.1M  & 8 & 6 & 656 & 27823 \\
\emph{Guns}       & 65743     & 167  &  187  & 65393 & 2.5M  & 98 & 115 & 56702 & 16608\\
\emph{Evolution}      & 84788    & 342  &  1334  & 83113 & 1.99M  & 391 & 135 & 15601 & 58720 \\
\emph{Racism}      & 129963   &  1024 & 1022	  & 127953 & 4.8M &  746 & 560 & 74354 & 58195\\
\emph{LGBT}      & 150563   &  459 & 640  & 149479 & 4.6M & 195 & 143 & 92975 & 81706\\
\end{tabular}}
\caption{Networks' statistics.}
\label{tab:info}
\end{table}

\subsubsection{\textbf{Partitions}}\label{topic_size}
In \cref{tab:info},\footnote{We add to the set $\onehop$ the articles assigned to both partitions.
The size of such intersections is: 2 (\emph{abortion}), 3 (\emph{cannabis}), 2 (\emph{evolution}), 1 (\emph{guns}), 5 (\emph{LGBT}), 7 (\emph{racism}). 
Because we do not remove these articles, they act as bridges connecting
$\p$ and $\pbar$ in sessions longer than one click.}
we observe that the size of $\p$ and $\pbar$ differs substantially, for all the topics but \emph{racism} and \emph{guns}. 
The disproportionate number of articles does not imply an unbalance in
content representation, but it can affect the partition's exposure within the entire Wikipedia network. 
The sizes of $\p$ and $\pbar$ are not linear in the number of edges across partitions. 
For instance, although the nodes in \emph{pro-life} are twice as many as those in \emph{pro-choice}, the links pointing to \emph{pro-choice} are 36\% more than those pointing to \emph{pro-life}.
This happens, with different magnitude, also for \emph{guns} and \emph{LGBT}. 

\subsubsection{\textbf{Hyperlinks across partitions}}\label{hyperlinks} 
\begin{figure}[t]
\begin{center}
\includegraphics[width=1\columnwidth]{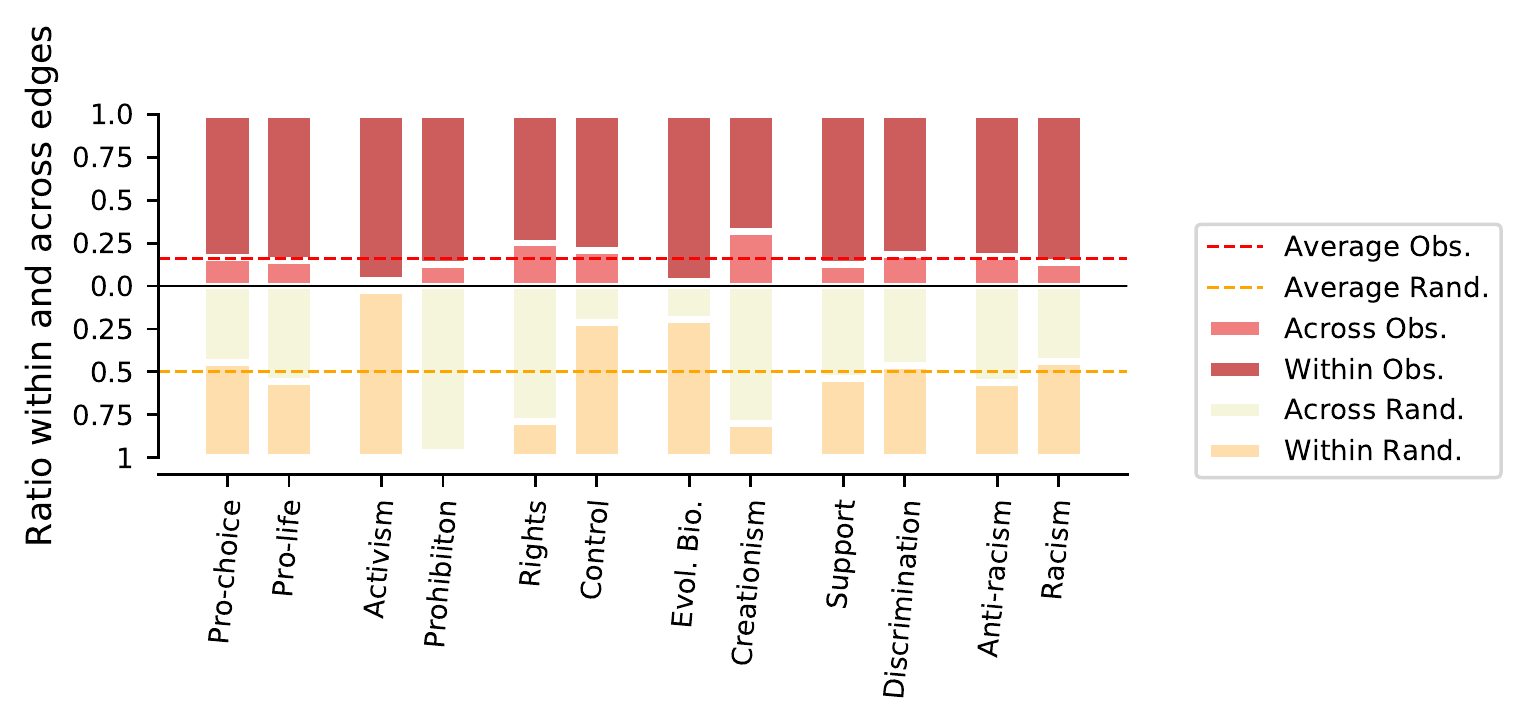}
\end{center}\vspace{-.5em}
\caption{Percentage of edges across and within partitions in each
\topinet (red) and a random graph with the same degree distribution
(orange). Topics in order are: \emph{abortion}, \emph{cannabis},
\emph{guns}, \emph{evolution}, \emph{racism}, and \emph{LGBT}.}
\label{fig:ration_edges}
\end{figure}

The direct exposure of users in $\p$ to pages in $\pbar$, depends on the
number of links connecting the two partitions.\footnote{We note that the
number of edges across \emph{cannabis}'s partitions is low,
nevertheless we keep the topic because on sessions longer that 1 click there are other paths connecting the partitions.}
To study their connectivity, we compare the portion of links in pages of $\p$ pointing to $\p$ and $\pbar$, with the same quantities expected on a random graph with the same degree sequence. 
In~\cref{fig:ration_edges}, we observe that most of the hyperlinks point to pages of the same partition.
On average fewer than 25\% of links point toward the opposing partition, which is against the 50\% expected on a random graph. 
The differences between the real and expected number of hyperlinks highlight that (1) links are, obviously, not randomly placed, (2) the strength of connections within and between partitions is skewed w.r.t. the distribution of edges conditioned on the number of nodes and their degree.
Furthermore, we speculate that the higher number of hyperlinks directed to pages of the same partition is due to the intrinsic clustered nature of Wikipedia~\cite{brandes2009network,lizorkin2009analysis}. 

\begin{table}[t]
\centering
\resizebox{1\columnwidth}{!}{
\begin{tabular}{lcccc}
\toprule
\textbf{p-values} & \textbf{$\onehop \rightarrow \p$ vs. $\onehop \rightarrow \pbar$}  & \textbf{Incoming higher avg.}                  & \textbf{$\p \rightarrow \onehop$ vs. $\pbar \rightarrow \onehop$}                      & \textbf{Outgoing higher avg.} \\ \midrule
\textit{Abortion}                     & $8.4\cdot 10^{-2}$  & -   & $9.6\cdot 10^{-1}$                              & -                             \\
\textit{Cannabis}                     & $6.5\cdot 10^{-8}$(**) & Activism   & $1.6\cdot 10^{-13}$(**)                     & Activism                      \\
\textit{Guns}                         & $1.3\cdot 10^{-4}$(**)  & Control  & $5.1\cdot 10^{-2}$                         & -                             \\
\textit{Evolution}                    & $7.2\cdot 10^{-3}$(**) & Creationism  & $4.9\cdot 10^{-5}$(**)                    & Creationism                   \\
\textit{Racism}                       & $6.2\cdot 10^{-6}$(**) & Anti-racism & $3.3\cdot 10^{-7}$(**)                     & Anti-racism                   \\
\textit{LGBT}                         & $1.4\cdot 10^{-2}$(**)  & Discrimination & $1.4\cdot 10^{-5}$(**)                 & Discrimination                \\ \bottomrule
\end{tabular}}
\caption{We report the p-values of t-tests ($\alpha=0.05$) on (1) the
number of links from $\onehop$ to $\p$ and $\pbar$ (first column), and
(2) the number of links to $\onehop$ from $\p$ and $\pbar$ (third
column). In the second and fourth columns we indicate which
partition is significantly more connected to the rest of Wikipedia.
Statistics are computed after bootstrapping the distributions of
from $\onehop$ to $\p$ and $\pbar$ and vice versa.}
\label{tab:pvalues}
\end{table}

\subsubsection{\textbf{Topic connectivity to the rest of Wikipedia}}\label{sec:nodes_rest_wiki}
We briefly investigate the connectivity between $\p$ (resp. $\pbar$)
with the rest of pages connected to it (i.e., $\onehop$). 
In~\cref{tab:pvalues}, we observe that for all the topics, but
\emph{abortion}, the average number of links coming from articles in
$\onehop$ and pointing to 
articles in $\onehop$ is significantly higher for one of the two
partitions. 

\subsubsection{\textbf{Distribution of across-partition links}}\label{sec:nodes_edges}
\begin{figure}[t]
\begin{center}
\includegraphics[width=1\columnwidth]{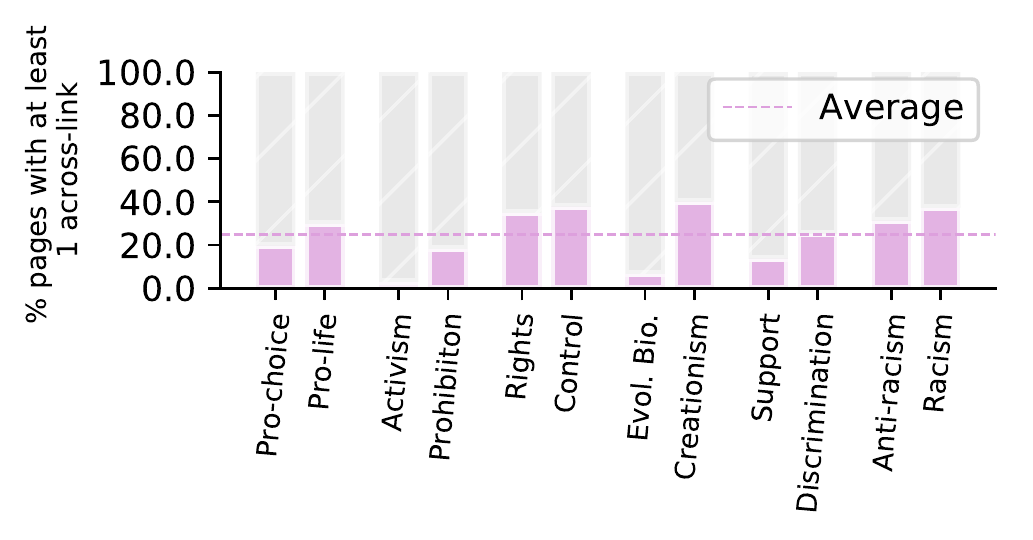}
\end{center}\vspace{-.5em}
\caption{Percentage of articles in $\p$ connected to $\pbar$. Topics are: \emph{abortion}, \emph{cannabis}, \emph{guns}, \emph{evolution}, \emph{racism} and \emph{LGBT}.}
\label{fig:across_link}
\end{figure}

If across-partition links are uniformly placed within articles of a
partition, users starting from an arbitrary node in the partition have the chance to visit pages about another branch of the topic. 
However, we observe that in our networks only a small subset of the
pages expose their visitors to another branch of the topic:
\cref{fig:across_link} shows that the average percentage of pages
connecting to the other partition is 25\% (average of 1.8 links per
    page), thus most of the nodes are not connected to the other
partition.

\subsubsection{\textbf{Weight distribution of across-partition links}}\label{sec:weights_edges}
The likelihood of traversing a link connected to the other side is conditioned on the number of links in a page.
\cref{fig:weight_link} shows that, for each topic, there is one
partition whose average probability of traversing an across-partition
hyperlink is statistically higher than the other partition (according to \emph{t-tests} with $\alpha=0.05$). 
For instance, the average chance to go from \emph{creationism} to \emph{evolutionary biology} are significantly lower than moving in the opposite direction.

%% file: metrics.tex
\section{Metrics}\label{sec:metrics}

In this section we introduce the metrics to quantify the exposure to
diversity, accounting for (1) the across-partition edges distribution
over nodes, (2) the likelihood of traversing a link toward the other
partition, and (3) the average exposure to diversity of all pages in a partition, considering navigation sessions of at least one click.

\subsection{Model of Readers' Behavior}\label{sec:models}

To comprehensively measure how much the network topology exposes users
to diversity, we should consider both the graph topology and how readers navigate the network. 
Indeed, the \divexp\ might vary for users who behave
differently in terms of navigation session length and next-link choices. 
So far, there are no models that generalize the navigation behavior of Wikipedia users. 
Thus, on top of previous findings~\cite{dimitrov2016visual,dimitrov2017success}, in~\cref{sec:click_model,sec:nav}, we define a parametric model that simulates a wide range of users' navigation sessions by embedding different behaviors. 
We emphasize that the scope of this model is not to perfectly replicate users' behavior on Wikipedia. Rather, we want to see how users simulated from a reasonable and general model get exposed to diverse information.

\input{figs/boxplot}

\subsubsection{\textbf{Model Clicks Within Pages (CwP)}}\label{sec:click_model}

When readers visit a page, they have the possibility of clicking any
link. However, according to the information needs they want to satisfy,
  each of the links may have a different
  click-probability~\cite{singer2017we}. We characterize the probability
  of ``clicking a link $j$ within an article $i$'' in three ways.
  First, let $i$ be an article in $V$ and
$j \in \outedges{i}$, where $\outedges{i}$ is the the set of pages to
which $i$ has a link. We define  $\mathit{pos}(j|i)$ as the rank of $j$
among all links in $i$, and $r(j|i)=|\outedges{i}| - \mathit{pos}(j|i)$, such that a higher value indicates a higher ranking position. 
We consider links in the infoboxes as at the top of the article, according to results in~\cite{dimitrov2016visual,dimitrov2017success}.
Moreover, we introduce $\tanh{x} =
\frac{e^{2x}-1}{e^{2x} + 1}$, which we use to transform ranking
positions to values between 0 and 1, such that links at the top of the
page are assigned similar scores. For instance if two links are adjacent
in a line, likely their probability of being clicked is similar.

We embed \emph{clicks within pages} models (CwP) into $G$ by setting its transition matrix $M$ in one of the following modes:
\begin{enumerate}
    \item  $M^u$ (\emph{Uniform}), whose entry
    $m(i,j)=\frac{1}{|N_{out}(i)|}$  mimics readers who click each link uniformly at random; 
    \item $M^p$ (\emph{Position}), whose entry
    $m(i,j)=\frac{\tanh{r(j|i)}}{\sum_{j \in N_{out}(i)}\tanh{r(j|i)}}$
    captures readers who click with higher probability links appearing
    on the top of the page. This model is based on previous
    works showing that the links' position is a good predictor to
    determine its success~\cite{dimitrov2017makes,lamprecht2017structure};
    \item $M^c$ (\emph{Clicks}), whose entry
    $m(i,j)=\frac{c_{i,j}}{\sum_{j \in N_{out}(i)} c_{i,j}}$ is the observed probability that users in $i$ will click the link
    toward $j$. The quantity $c_{i,j}$ counts how many times on average real users clicked the hyperlink from page $i$ to $j$, from August 2019 to September 2020.\footnote{Wikipedia's clickstream data is publicly available and preserves users' privacy~\cite{wulczyn2016growing,dimitrov2019clicks}. Data description at \url{Research:Wikipedia\_clickstream}.}
    For the links never clicked, we set $c_{ij}=10$, the minimum number
    of times that the link must be clicked to be included in the
    dataset~\cite{wulczyn2017clickstream}. This smoothing factor allows one to assign a positive weight to links rarely clicked.
\end{enumerate}

\input{figs/behaviour_models}


\subsubsection{\textbf{Readers Navigation Model}}\label{sec:nav}
To characterize the users' sessions, we define a stochastic process with
$\lvert V\rvert$ states, which, for each click, approximates the probability of reaching any of the articles starting at random from $p \in \p$ (or from $\pbar$). 
We consider the process $\{X^\ell; \ell=0,1,\dots\, L\}$, on the set of
nodes $V$ induced by the transition matrix $M$ with starting state $X^0$ selected from the probability distribution $\boldsymbol{\pi}_P^0=\{(\pi_P)_i\}\in\mathbb{R}^{1\times n}$ over $V$. 
Assuming that the user session length (the number of clicks) is finite,
we evaluate the process on a finite number of steps $L$.
Thus, $\textbf{Pr}(X^\ell=j) = (\pi_P^\ell)_j$, where the
(row) vector $\boldsymbol{\pi}_P^{\ell}$ is given by the following
variation of the Personalized Random Walk with Restart (RWR).

\begin{definition}[Navigation Model] \textit{Let $M_0$ be the transition matrix embedding a
click-within-pages model, $\boldsymbol{\pi}_P^0$ the distribution of
the starting state over $P$, and $\alpha \in [0,1]$ the restart parameter.
We have}
\begin{equation}
    \boldsymbol{\pi}_P^{1}  = \boldsymbol{\pi}_P^{0} \cdot M_0
\end{equation}
and, for $\ell\ge1$,
\begin{equation}
    \boldsymbol{\pi}_P^{\ell+1}  = (1-\alpha)\boldsymbol{\pi}_P^{\ell} \cdot M_\ell + \alpha(\boldsymbol{\pi}_P^{0} \cdot M_\ell),
\end{equation}
\textit{where $M_\ell=\mathit{\mathit{norm}}((D(M_{\ell-1})^T)^T)$ and
$D=\mathit{diag}\left(\mathbf{1}+\boldsymbol{\pi}_P^{\ell-1}\right)^{-1}$.
The operator $\mathit{norm}(M)$ transforms matrix $M$ into a right-stochastic
matrix by normalizing each row independently such that it sums to~1.}
\end{definition}

\input{figs/local_exposure}

This process is a variation of the standard random-surfer model that differs for the update of the transition matrix at each step.
The vector $\boldsymbol{\pi}_P^{\ell}$ represents the likelihood that each node is reached at step $\ell$ if the session starts uniformly at random from a node in $P$.
Assuming that readers do not click multiple times the same link within a session, we desire to deflate the probability of reaching nodes that, at step $\ell+1$, have already been visited with high probability.
We achieve this by dividing the rows of $M$ by the vector of
probabilities $\boldsymbol{\pi}_P^{\ell}$+1, where 1 is a smoothing
factor, and then normalize the matrix to obtain the updated stochastic matrix to use in the next iteration. 
Looking deeper into the model:

\begin{itemize}
    \item For $\alpha=0$ (\cref{fig:dfs}), the readers' clicks depend only on the CwP model. In this case, especially if related articles are not densely connected, the exploration can quickly lead to articles less related to the starting page.   

    \item For $\alpha=1$ (\cref{fig:bfs}), readers locally explore articles likely semantically related to each
    other~\cite{wikipediaManualStyle}, and the model emulates a
    \emph{star-like} behavior, which consists in sequentially opening links from the starting page.

    \item For $0<\alpha<1$ (\cref{fig:dfs_star}), the readers' choices
    depend on the CwP model and, occasionally, they go back to the
    initial page. The more $\alpha$ is close to 1 the more users show a star-like behavior.
    The closer $\alpha$ is to 0 the more users navigate navigate in a more Depth First Search-oriented fashion.
    The model emulates (1) readers who sequentially explore articles and then jump back to the starting page, or (2) readers keeping open multiple paths. 
\end{itemize}

Wikipedia does not have a button that allows readers to go back to
the previous page.
Thus, to \emph{jump back} consists of clicking the browser's back button, until the session starting page. 
The restart parameter indirectly embeds the back button, which for the absence of \emph{back-links} on Wikipedia does not appear in the graph.

\subsection{Quantification of Exposure to Diverse Information}\label{sec:exposure}

The \textit{exposure to diverse information} aims to quantify how much the network structure allows readers to reach one, or multiple sets of articles, depending on their behavior.
It is built upon both the CwP and the \emph{navigation} models, and its application generalizes to arbitrary sets of nodes in a graph.

\begin{definition}[\udivexp\ (\Adivexp)]\label{def:divexp}
\textit{Given two sets of pages $\p, \pbar$ in $V$, let
$\boldsymbol{\pi}_{\p}^{\ell}$ be the vector indicating the
probability distribution of reaching any node in $V$ at step $\ell$ ($\ell\ge1$) starting from a random page in $\p$. We say
that the exposure of $\p$ to $\pbar$ is}
\begin{equation}
    \exposure{\p}{\pbar}{\ell} = \sum_{j \in \pbar} \textbf{Pr}(X^\ell = j) = \sum_{j \in \pbar} (\pi_P^\ell)_j
\end{equation}
\textit{and it describes the probability that a reader in $\p$ reaches an arbitrary
node in $\pbar$ at the $\ell$th click.}
\end{definition}


\input{figs/dynamic_exposure}

\Cref{def:divexp} can be extended to multiple sets. 
Assume that we want to measure how much the set $\p$ is exposed to three sets of nodes, $Q, Z$, and $L$. 
The total exposure to the three sets is the \Adivexp computed setting $\pbar= Q \cup Z \cup L$. 
Otherwise, if we want to have the \Adivexp w.r.t. to each set, namely,
$e_{\p \rightarrow Q}, e_{\p \rightarrow Z}$, and $e_{\p \rightarrow L}$, we take $\boldsymbol{\pi}_{\p}^{\ell}$ and sum up the probabilities of the nodes within each set. 

To quantify the extent to which the \divexp is balanced across $\p$ and $\pbar$, we introduce the \emph{\lmutual}.  

\begin{definition}[\mutual\ (\Amutual)] \textit{Let $e_{\p \rightarrow \pbar}^\ell$ and
$e_{\pbar \rightarrow \p}^\ell$ be the \divexp of sets $\p$ and $\pbar$.
The mutual exposure between the sets is}
\begin{equation}
    \epsilon^\ell  = \frac{\min \{e_{\p \rightarrow \pbar}^\ell, e_{\pbar \rightarrow \p}^\ell\}}{\max\{e_{\p \rightarrow \pbar}^\ell, e_{\pbar \rightarrow \p}^\ell\}} \in [0,1].
\end{equation}
\textit{If either $e_{\p \rightarrow \pbar}^\ell$ or $e_{\pbar \rightarrow \p}^\ell$ is 0, then $\epsilon=0$.}
\end{definition}

The closer $\epsilon$ is to 1, the more balanced are the probabilities of moving from one set to the other are. 
Thus, the network topology does not favor connections from one set to the other.
Otherwise, the network structure tends to favor the navigation from one partition toward the other.
With this view, if the network structure facilitates to move from one of the sets to the other, we may say that the network topology is \emph{biased} toward a direction.
Thus, \Amutual\ is a measure of the network's bias w.r.t. two sets of nodes, at each click of a session. 
If the sizes of sets $\p$ and $\pbar$ are unbalanced, we obtain higher probabilities for large partitions. 
To take the sizes into account, we introduce a strategy to compute the adjusted-\Adivexp. Given the two partitions $\p$ and $\pbar$, we set a sample size equal to $z = \min \{|\p|,|\pbar|\}$. From the two sets we sample with replacement $\p_i'$ and $\pbar_i'$ of size $z$, respectively from the initial partitions. Hence, we bootstrap $\exposure{\p'}{\pbar'}{\ell}$ estimating the value of the adjusted-\Adivexp.

%% file: figs/boxplot.tex
\begin{figure}[t]
\begin{center}
\includegraphics[width=1\columnwidth]{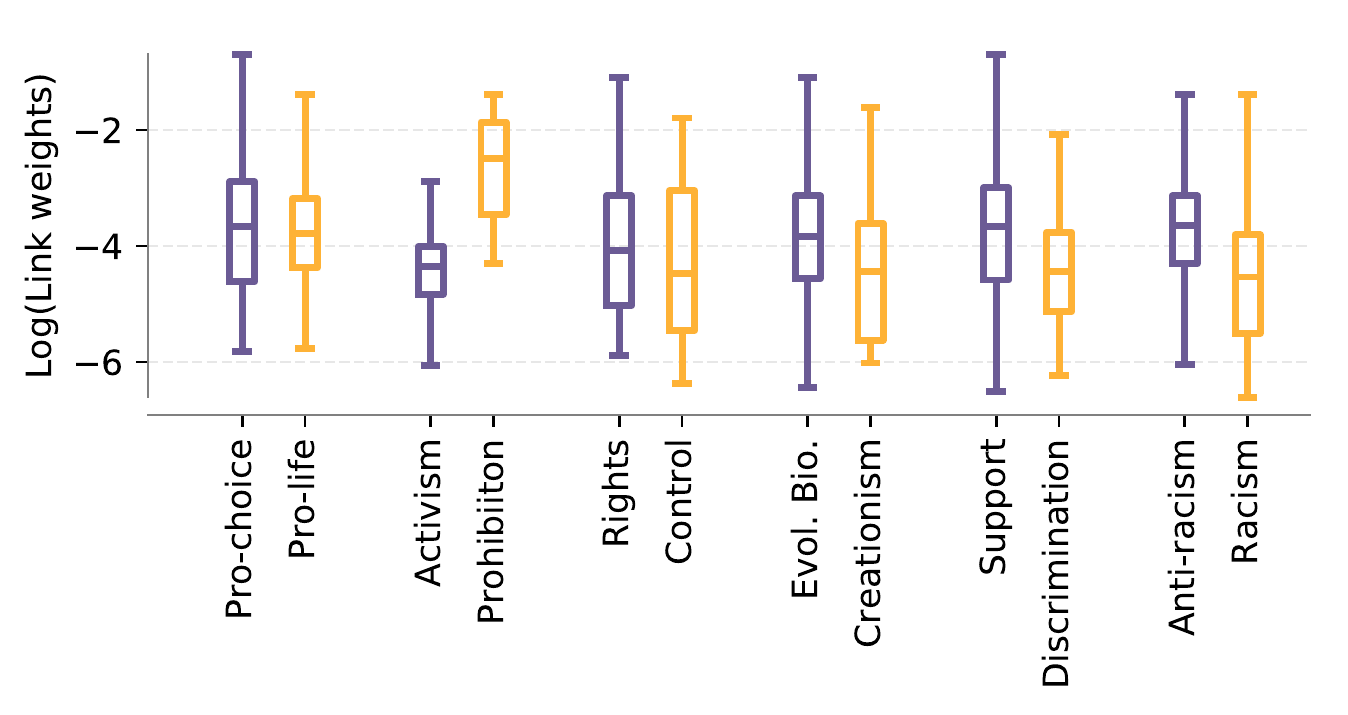}
\end{center}\vspace{-.5em}
\caption{Distributions of across-partition links weights. Topics are: \emph{abortion}, \emph{cannabis}, \emph{guns}, \emph{evolution}, \emph{racism} and \emph{LGBT}.}
\label{fig:weight_link}
\end{figure}

%% file: figs/behaviour_models.tex
\begin{figure}[t]
\begin{center}
\subfigure[Random Navigation ($\alpha=0$)]{\label{fig:dfs}
\includegraphics[width=0.12\textwidth]{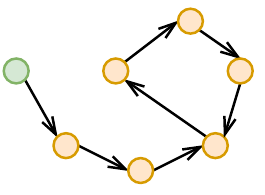}}
\hspace*{0.1em}
\subfigure[Star-like \mbox{($\alpha=1$)}]{\label{fig:bfs}
\includegraphics[width=0.13\textwidth]{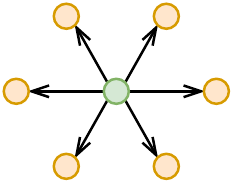}}
\hspace*{0.1em}
\subfigure[Star-like Random Navigation ($0<\alpha<1$)]{\label{fig:dfs_star}
\includegraphics[width=0.19\textwidth]{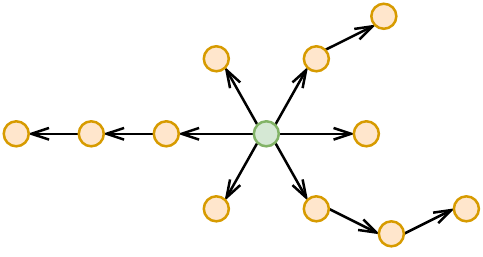}}
\end{center}\vspace{-.5em}
\caption{Navigation model for different $\alpha$.}
\label{fig:navmodel}
\end{figure}

%% file: figs/local_exposure.tex
\begin{figure*}[t]
\begin{center}
\subfigure{\label{fig:exdin:boot}
\includegraphics[width=1\textwidth]{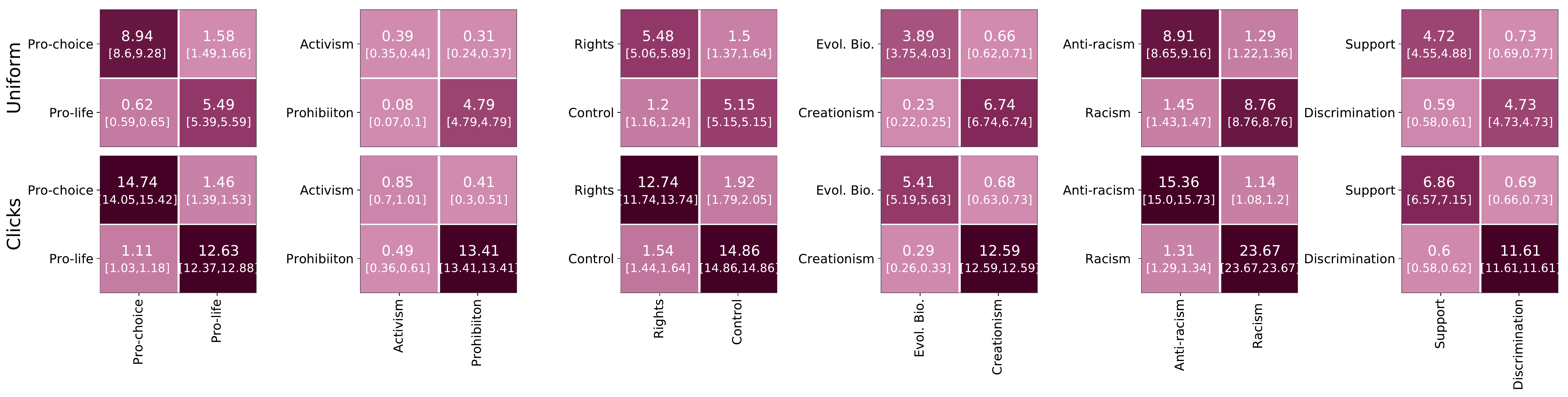}
}
\end{center}\vspace{-.5em}
\caption{\emph{Local exposure to diversity}. Each plot shows the adjusted-ExDIN (\%) across
partitions. On the main diagonal we have flows within the same partition, e.g., $\p \rightarrow \p$. The off-diagonal reports the probability of moving across sides, e.g., $\p \rightarrow \pbar$. The
$y$-axis indicates the source and the $x$-axis is the destination. To
each row corresponds the \divexp\ computed for different CwP. Darker
colors indicate higher probability of being in the
corresponding square in one click. The values in the brackets are the
90\% confidence intervals. Topics are: \emph{abortion, cannabis, guns,
  evolution, racism}, and \emph{LGBT}.}
\label{fig:exdin}
\end{figure*}

%% file: figs/dynamic_exposure.tex
\begin{figure*}[ht]
\begin{center}
\includegraphics[width=1\textwidth]{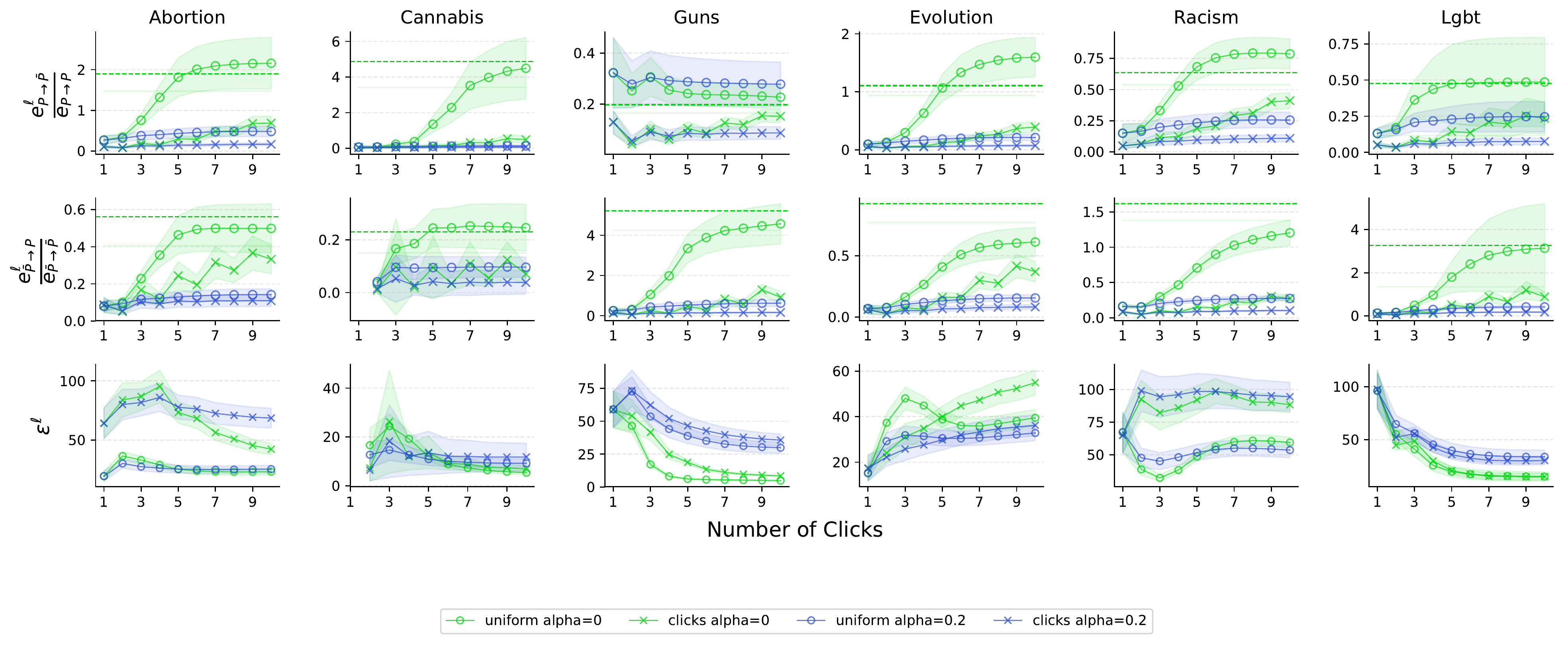}
\end{center}\vspace{-.5em}
\caption{\emph{(Adjusted) Dynamic Exposure to Diversity for sessions of 1 to 10 clicks}. The $x$-axis indicates the number of clicks. In the
first two rows, the $y$-axis is the ratio of the probability of
moving from $\p$ to $\pbar$ and $\p$, and
vice versa. Values lower than 1 indicate that remaining in the
\emph{knowledge bubble} is more likely than visiting pages of
diverse content. Values close to zero quantify the chances of
staying in a \emph{knowledge bubble}. The $y$-axis of the third row
indicates the mutual exposure to diversity (\%). Values closer to 0 show
that the current network topology might be biased toward the partition
$\p$ or $\pbar$. Colors indicate the navigation model (set by $\alpha$).
Shapes encode the CwP model. The bands indicate the standard deviation. The
dashed lines represent the convergence values for infinite sessions.}
\label{fig:mutualexdin}
\end{figure*}

%% file: exposure.tex
\section{Exposure to Diverse Viewpoints}\label{sec:exposure_main}
\subsection{Local Exposure to Diversity}\label{sec:exposure_to_diversity}

The \emph{local} exposure to diversity is the possibility of accessing articles about another branch of the topic within one click.
We measure it using \divexp, setting $\ell=1$, which describes the static connectivity among partitions accounting for users' choices and the network topology. 

In~\cref{fig:exdin}, the local exposure to diversity on each \topinet
shows the following:\footnote{We omit the plots showing result when using
the model embedding the links' position. In general, there are not
significant differences compared to the uniform model. 
For some topics, such as \emph{guns}, the links’ position plays a more significant role, worsening the user exposure to diverse information.} 
When considering only the graph topology (i.e., $M^u$):
(1) The networks' topology facilitates users to remain in a \emph{knowledge bubble} (i.e., same partition) hindering the exploration of the topic's diverse stances. 
For every topic the probability of visiting pages of the opposing partition is on average 12 times lower than staying in the initial one. 
(2) One of the two partitions induces higher chances of remaining within the same bubble. 
On average, among all topics but \emph{cannabis}, one of the two partitions has 2 times more chances of keeping users within its articles. 

After embedding users' past behavior (i.e., $M^u$):
(3) The probability that readers keep visiting pages about the same topic click-after-click (i.e., remaining within a \emph{knowledge bubble}) is higher than when users click links uniformly at random (i.e., $M^u$).
Indeed, on average, users have 1.5 more chances of moving within $\topicpages$. 
The probability increases significantly, suggesting to start a discussion on the importance of exposing users to diverse information. 
(4) The likelihood of exploring diverse content slightly increases, showing that real users clicked pages of the opposing partitions more than what described by the uniform model.
On average, users have 1.6 times more chances of moving to the opposing sides. %
(5) The discrepancy between the probability of moving within and outside the initial partition is on average even 2.6 times larger than when using $M^u$.

From past users' clicks we observe the preference of navigating across pages of the same partition. 
So far, users' behavior has always been justified by their information needs. 
From the observations (1) and (2), we know that the network's topology potentially favors the visit of pages of same standing. 
Is it possible that users' next-click choices are influenced by the hyperlinks network? 


\subsection{Dynamic Exposure to Diversity}\label{sec:dynexposure}

We expand the analysis to sessions longer than one click to study the users' dynamic exposure to diversity. 
In~\cref{fig:mutualexdin}, we observe that:
(1) Within a few steps (4--5 clicks), users are more exposed to the pages of the same partition (i.e., the ratio of exposures is smaller than 0). 
(2) The current structure of the network favors users starting from any random page to reach one partition more easily than the other. 
Can we consider it a bias of the network? 
Is it fair that one side of a polarizing topic is less reachable than the opposite from any random node in the graph?
(3) If users navigate according to a star-like random navigation model, the ratio between moving outside and within the partition stays steady or slightly increases.
(4) All \topinets are topologically biased; indeed none of them does the
network provides an equal exposure across partitions (i.e., \lmutual is lower than 100\%).
(5) The local bias is smaller than the overall bias of the network. In general, the \lmutual\ decreases for longer sessions. 
(6) The general topology of the graph makes pages related to
\emph{liberal} standings more accessible.\footnote{We omit the star-like
model ($\alpha=0$), because its value is steady around the value for
$\ell=1$, and we omit the position CwP model because it shows trends similar to uniform.}

The lack of mutual exposure might depend on many factors such as
\emph{underlinked} articles~\cite{wikipediaManualStyle} or missing words to attach links. An in-depth investigation of this condition may be an
interesting future work.

\subsection{Factors Related to Exposure to Diversity}


\subsubsection{ExDIN and homophily} 
We measure the Pearson correlation between the homophily\footnote{We
use the EI Homophily index~\cite{Krackhardt1988InformalNA}: $EI(v \in \p)=\frac{|\mathit{ext}_{\p}|-|\mathit{int}_{\p}|}{|\mathit{ext}_{\p}|+|\mathit{int}_{\p}|}$,
where $\mathit{ext}_{\p}$ is the set of edges from $\p$ to the
rest of the network, and $\mathit{int}_{\p}$ is the number of edges pointing to $\p$.} of article $i$ and the users' exposure to diversity starting a session in $i$.
Locally, starting from a page with high homophily decreases users' probabilities of being exposed to diverse content. 
Indeed, we observe a negative correlation for 1--3 clicks sessions (avg.
$-0.45$), and an average correlation close to $0$, for longer sessions. 

\subsubsection{ExDIN and centrality}\label{sec:centrality} 
We measure the Pearson correlation between the degree centrality of article $i$ and the users' exposure to diversity starting a session in $i$. 
The number of incoming links of an initial page does not play a role in determining the exposure to diversity, especially within a few clicks (in-degree correlation 0.07).
On the other hand, within 1--3 clicks, the lower out-degree mildly increases the users exposure to diversity (average -0.25). Under the uniform model, it is because the links' transition probabilities of pages with a few links are larger than pages with a lot of links. 

%% file: concl.tex
\section{Discussions}\label{sec:discussion}

In this work, we look for the first time at Wikipedia's hyperlink structure to measure its influence on users' exposure to diverse information. 
By employing two Wikipedia-tailored metrics, we quantify the likelihood of visiting pages representing different aspects of a topic throughout a navigation session. 
Our findings indicate that the current network topology often limits
exposure to diverse information and incentivizes users to remain in
\emph{knowledge bubbles}. 

The ultimate goal of this work is to draw attention and initiate a
discussion about the importance of evaluating the hyperlink structure as
part of Wikipedia's goal to provide a natural point of view presentation, even for polarizing subjects.
Our observations raise a number of interesting questions for the Wikipedia community. 
As an example,  consider a page about an \emph{anti-abortion organization}. 
It seems natural that this page has more hyperlinks to pages related to anti-abortion subjects than to pages related to abortion rights. 
This is reasonable and aligns with the current purpose of Wikipedia's
internal links, but is it still reasonable, and conforms with the goal
of a natural point of view?
Similarly, we observe that in the directed hyperlink graph, it is often more likely to reach an article about B starting from an article about A, than reaching an article about A starting with an article about B, when A and B represent two aspects of a topics.
Again, some imbalance is reasonable, but it can keep users locked in an information bubble. 
How do we distinguish between the two cases? 

We expect ours and future findings to motivate work on editors' support tools for contextualizing pages within their neighborhood in the hyperlink network, and suggest hyperlink modifications to improve access to diverse content.

\section*{Acknowledgements}
The project is supported by DARPA LwLL program,
by the ERC Advanced Grant 788893 AMDROMA, the EC H2020RIA project ``SoBigData++'' (871042), and the MIUR PRIN
project ALGADIMAR.
The views and conclusions contained herein are those
of the authors and should not be interpreted as representing the official policies or endorsements, either expressed or implied, of DARPA or the U.S. Government.

%% file: main.bbl
\begin{thebibliography}{10}
\providecommand{\url}[1]{#1}
\csname url@samestyle\endcsname
\providecommand{\newblock}{\relax}
\providecommand{\bibinfo}[2]{#2}
\providecommand{\BIBentrySTDinterwordspacing}{\spaceskip=0pt\relax}
\providecommand{\BIBentryALTinterwordstretchfactor}{4}
\providecommand{\BIBentryALTinterwordspacing}{\spaceskip=\fontdimen2\font plus
\BIBentryALTinterwordstretchfactor\fontdimen3\font minus
  \fontdimen4\font\relax}
\providecommand{\BIBforeignlanguage}[2]{{%
\expandafter\ifx\csname l@#1\endcsname\relax
\typeout{** WARNING: IEEEtran.bst: No hyphenation pattern has been}%
\typeout{** loaded for the language `#1'. Using the pattern for}%
\typeout{** the default language instead.}%
\else
\language=\csname l@#1\endcsname
\fi
#2}}
\providecommand{\BIBdecl}{\relax}
\BIBdecl

\bibitem{wikipediaManualStyle}
Wikipedia, ``Linking,'' in \emph{Wikipedia:Manual of style/linking}.

\bibitem{keegan2011hot}
B.~Keegan, D.~Gergle, and N.~Contractor, ``Hot off the wiki: dynamics,
  practices, and structures in wikipedia's coverage of the t{\=o}hoku
  catastrophes,'' in \emph{Proc. of the 7th international symposium on Wikis
  and open collaboration}, 2011.

\bibitem{piscopo2019we}
A.~Piscopo and E.~Simperl, ``What we talk about when we talk about wikidata
  quality: a literature survey,'' in \emph{Proc. of the 15th International
  Symposium on Open Collaboration}, 2019.

\bibitem{wikipediaNPOV}
Wikipedia, ``Neutral point of view,'' in
  \emph{Wikipedia:Neutral\_point\_of\_view}.

\bibitem{beschastnikh2008wikipedian}
I.~Beschastnikh, T.~Kriplean, and D.~W. McDonald, ``Wikipedian self-sovernance
  in action: motivating the policy lens.'' in \emph{ICWSM}, 2008.

\bibitem{forte2009decentralization}
A.~Forte, V.~Larco, and A.~Bruckman, ``Decentralization in wikipedia
  governance,'' \emph{Journal of Management Information Systems}, 2009.

\bibitem{singer2017we}
P.~Singer, F.~Lemmerich, R.~West, L.~Zia, E.~Wulczyn, M.~Strohmaier, and
  J.~Leskovec, ``Why we read wikipedia,'' in \emph{Proc. of the 26th
  International Conference on World Wide Web}, 2017.

\bibitem{wikipediaStats}
Wikipedia, ``Statistics,'' in \emph{Wikipedia:Statistics}.

\bibitem{ribeiro2020auditing}
M.~H. Ribeiro, R.~Ottoni, R.~West, V.~A. Almeida, and W.~Meira~Jr, ``Auditing
  radicalization pathways on youtube,'' in \emph{Proc. of FAccT 2020}.

\bibitem{redi2019citation}
M.~Redi, B.~Fetahu, J.~Morgan, and D.~Taraborelli, ``Citation needed: A
  taxonomy and algorithmic assessment of wikipedia's verifiability,'' in
  \emph{The World Wide Web Conference}, 2019.

\bibitem{fetahu2016finding}
B.~Fetahu, K.~Markert, W.~Nejdl, and A.~Anand, ``Finding news citations for
  wikipedia,'' in \emph{Proc. of the 25th ACM International on Conference on
  Information and Knowledge Management}, 2016.

\bibitem{piccardi2018structuring}
T.~Piccardi, M.~Catasta, L.~Zia, and R.~West, ``Structuring wikipedia articles
  with section recommendations,'' in \emph{The 41st International ACM SIGIR
  Conference}, 2018.

\bibitem{oaxes}
S.~Kumar, R.~West, and J.~Leskovec, ``Disinformation on the web: Impact,
  characteristics, and detection of wikipedia hoaxes,'' in \emph{Proc. of the
  25th International Conference on World Wide Web}, 2016.

\bibitem{paranjape2016improving}
A.~Paranjape, R.~West, L.~Zia, and J.~Leskovec, ``Improving website hyperlink
  structure using server logs,'' in \emph{Proc. of the Ninth ACM International
  Conference on Web Search and Data Mining}, 2016.

\bibitem{wulczyn2016growing}
E.~Wulczyn, R.~West, L.~Zia, and J.~Leskovec, ``Growing wikipedia across
  languages via recommendation,'' in \emph{Proc. of the 25th International
  Conference on World Wide Web}, 2016.

\bibitem{helic2013models}
D.~Helic, M.~Strohmaier, M.~Granitzer, and R.~Scherer, ``Models of human
  navigation in information networks based on decentralized search,'' in
  \emph{Proc. of the 24th ACM conference on hypertext and social media}, 2013.

\bibitem{gildersleve2018inspiration}
P.~Gildersleve and T.~Yasseri, ``Inspiration, captivation, and misdirection:
  Emergent properties in networks of online navigation,'' in
  \emph{International Workshop on Complex Networks}, 2018.

\bibitem{lamprecht2017structure}
D.~Lamprecht, K.~Lerman, D.~Helic, and M.~Strohmaier, ``How the structure of
  wikipedia articles influences user navigation,'' \emph{New Review of
  Hypermedia and Multimedia}, 2017.

\bibitem{singer2013human}
P.~Singer, T.~Niebler, M.~Strohmaier, and A.~Hotho, ``Computing semantic
  relatedness from human navigational paths: A case study on wikipedia,'' in
  \emph{International Journal on Semantic Web and Information Systems 9}, 2013.

\bibitem{west2012human}
R.~West and J.~Leskovec, ``Human wayfinding in information networks,'' in
  \emph{Proc. of the 21st international conference on World Wide Web}, 2012.

\bibitem{scaria2014last}
A.~T. Scaria, R.~M. Philip, R.~West, and J.~Leskovec, ``The last click: Why
  users give up information network navigation,'' in \emph{Proc. of the 7th ACM
  international conference on Web search and data mining}, 2014.

\bibitem{dallmann2016extracting}
A.~Dallmann, T.~Niebler, F.~Lemmerich, and A.~Hotho, ``Extracting semantics
  from random walks on wikipedia: Comparing learning and counting methods.'' in
  \emph{Wiki@ICWSM}, 2016.

\bibitem{koopmann2019right}
T.~Koopmann, A.~Dallmann, L.~Hettinger, T.~Niebler, and A.~Hotho, ``On the
  right track! analysing and predicting navigation success in wikipedia,'' in
  \emph{Proc. of the 30th ACM Conference on Hypertext and Social Media}, 2019.

\bibitem{dimitrov2016visual}
D.~Dimitrov, P.~Singer, F.~Lemmerich, and M.~Strohmaier, ``Visual positions of
  links and clicks on wikipedia,'' in \emph{Proc. of the 25th International
  Conference Companion on World Wide Web}, 2016.

\bibitem{dimitrov2017success}
------, ``What makes a link successful on wikipedia?'' in \emph{Proc. of WWW
  2017}.

\bibitem{blei2010probabilistic}
D.~Blei, L.~Carin, and D.~Dunson, ``Probabilistic topic models,'' \emph{IEEE
  signal processing magazine}, 2010.

\bibitem{shi2019wisdom}
F.~Shi, M.~Teplitskiy, E.~Duede, and J.~A. Evans, ``The wisdom of polarized
  crowds,'' \emph{Nature human behaviour}, 2019.

\bibitem{adamic2005political}
L.~A. Adamic and N.~Glance, ``The political blogosphere and the 2004 us
  election: divided they blog,'' in \emph{Proc. of the WWW-2005 Workshop on the
  Weblogging Ecosystem}, 2005.

\bibitem{CossardDFMKMPS20}
A.~Cossard, G.~De~Francisci~Morales, K.~Kalimeri, Y.~Mejova, D.~Paolotti, and
  M.~Starnini, ``Falling into the echo chamber: The {I}talian vaccination
  debate on {T}witter,'' in \emph{Proc. of the International AAAI Conference on
  Web and Social Media}, 2020.

\bibitem{conover2011political}
M.~D. Conover, J.~Ratkiewicz, M.~Francisco, B.~Gon{\c{c}}alves, F.~Menczer, and
  A.~Flammini, ``Political polarization on twitter,'' in \emph{Fifth
  international AAAI conference on weblogs and social media}, 2011.

\bibitem{flaxman2016filter}
S.~Flaxman, S.~Goel, and J.~M. Rao, ``Filter bubbles, echo chambers, and online
  news consumption,'' \emph{Public opinion quarterly}, 2016.

\bibitem{guerra2013measure}
P.~C. Guerra, W.~Meira~Jr, C.~Cardie, and R.~Kleinberg, ``A measure of
  polarization on social media networks based on community boundaries,'' in
  \emph{7th International AAAI Conference on Weblogs and Social Media}, 2013.

\bibitem{garimella2018quantifying}
K.~Garimella, G.~D.~F. Morales, A.~Gionis, and M.~Mathioudakis, ``Quantifying
  controversy on social media,'' \emph{ACM Transactions on Social Computing},
  2018.

\bibitem{repbublik}
S.~Haddadan, C.~Menghini, M.~Riondato, and E.~Upfal, ``Repbublik: Reducing
  polarized bubble radius with link insertions,'' in \emph{Proc. of WSDM 2021}.

\bibitem{callahan2011cultural}
E.~S. Callahan and S.~C. Herring, ``Cultural bias in wikipedia content on
  famous persons,'' \emph{Journal of the American society for information
  science and technology}, 2011.

\bibitem{genderbias}
E.~Graells-Garrido, M.~Lalmas, and F.~Menczer, ``First women, second sex:
  Gender bias in wikipedia,'' in \emph{Proc. of the 26th ACM Conference on
  Hypertext \& Social Media}, 2015.

\bibitem{wagner2016women}
C.~Wagner, E.~Graells-Garrido, D.~Garcia, and F.~Menczer, ``Women through the
  glass ceiling: gender asymmetries in wikipedia,'' \emph{EPJ Data Science},
  2016.

\bibitem{wikipediaNamespace}
Wikipedia, ``Namespace,'' in \emph{Wikipedia:Namespace}.

\bibitem{wikipediaRedirect}
------, ``Redirect,'' in \emph{Wikipedia:Redirect}.

\bibitem{brandes2009network}
U.~Brandes, P.~Kenis, J.~Lerner, and D.~Van~Raaij, ``Network analysis of
  collaboration structure in wikipedia,'' in \emph{Proc. of the 18th
  international conference on World wide web}, 2009.

\bibitem{lizorkin2009analysis}
D.~Lizorkin, O.~Medelyan, and M.~Grineva, ``Analysis of community structure in
  wikipedia,'' in \emph{International conference on World wide web}, 2009.

\bibitem{dimitrov2017makes}
D.~Dimitrov, P.~Singer, F.~Lemmerich, and M.~Strohmaier, ``What makes a link
  successful on wikipedia?'' in \emph{Proc. of the 26th International
  Conference on World Wide Web}, 2017.

\bibitem{dimitrov2019clicks}
D.~Dimitrov and F.~Lemmerich, ``Democracy and difference: Different topic,
  different traffic: How search and navigation interplay on wikipedia,''
  \emph{The Journal of Web Science 6}, 2019.

\bibitem{wulczyn2017clickstream}
E.~Wulczyn and D.~Taraborelli, ``Wikipedia clickstream,''
  \emph{https://doi.org/10.6084/m9.figshare.1305770.v22}, 2017.

\bibitem{Krackhardt1988InformalNA}
D.~Krackhardt and R.~N. Stern, ``Informal networks and organizational crises:
  An experimental simulation,'' \emph{Social Psychology Quarterly}, 1988.

\end{thebibliography}
